# Grid enabled virtual screening against malaria

N. Jacq[1+2], J. Salzemann[1], F. Jacq[1], Y. Legré[1], E. Medernach[1], J. Montagnat[3], A. Maaß[4], M. Reichstadt[1], H. Schwichtenberg[4], M. Sridhar[5], V. Kasam[5], M. Zimmermann[5], M. Hofmann[5] and V. Breton[1]

[1] Laboratoire de Physique Corpusculaire, Université Blaise Pascal/IN2P3-CNRS UMR 6533, France

[2] Communication & Systèmes, France

[3] Informatique Signaux et Systèmes, Université de Nice Sophia Antipolis/CNRS UMR 6070, France

[4] Fraunhofer Institute for Algorithms and Scientific Computing, Department of Simulation Engineering, Germany

[5] Fraunhofer Institute for Algorithms and Scientific Computing, Department of Bioinformatics, Germany

Corresponding author: N. Jacq, jacq@clermont.in2p3.fr, LPC Campus des Cézeaux 24 av des Landais 63177 Aubière France, 33 4 73 40 53 24, 33 4 73 26 45 98

**Abstract**

WISDOM is an international initiative to enable a virtual screening pipeline on a grid infrastructure. Its first attempt was to deploy large scale in silico docking on a public grid infrastructure. Protein-ligand docking is about computing the binding energy of a protein target to a library of potential drugs using a scoring algorithm. Previous deployments were either limited to one cluster, to grids of clusters in the tightly protected environment of a pharmaceutical laboratory or to pervasive grids. The first large scale docking experiment ran on the EGEE grid production service from 11 July 2005 to 19 August 2005 against targets relevant to research on malaria and saw over 41 million compounds docked for the equivalent of 80 years of CPU time. Up to 1,700 computers were simultaneously used in 15 countries around the world. Issues related to the deployment and the monitoring of the *in silico* docking experiment as well as experience with grid operation and services are reported in the paper. The main problem encountered for such a large scale deployment was the grid infrastructure stability. Although the overall success rate was above 80%, a lot of monitoring and supervision was still required at the application level to resubmit the jobs that failed. But the experiment demonstrated how grid infrastructures have a tremendous capacity to mobilize very large CPU resources for well targeted goals during a significant period of time. This success leads to a second computing challenge targeting Avian Flu neuraminidase N1.

# 1. Introduction

## 1.1. Current challenges in high throughput screening

Screening is the first step of the drug discovery process once a molecular target has been identified. It is about selecting the compounds, i.e. the molecules, which could impact the target biochemical activity.

Advances in combinatorial chemistry have paved the way for synthesizing millions of different chemical compounds. Thus there are millions of chemical compounds available in the laboratories and recorded in 2D or 3D electronic databases, but it is nearly impossible and very expensive to screen such a high number of compounds in the experimental laboratories using high throughput screening techniques. Besides the heavy costs, the hit rate in high throughput screening is quite low; it is in the range of 1 per 100,000 compounds when screened on targets such as enzymes [1].

An alternative is high throughput virtual screening by molecular docking, a technique which can screen millions of compounds rapidly, reliably and cost effectively. Basically, protein-compound docking is about computing the binding energy of a protein target to a library of potential drugs using a scoring algorithm. The target is typically a protein (the "target") which plays a pivotal role in a pathological process, e.g. the biological cycles of a given pathogen (parasite, virus, bacteria…). Screening millions of chemical compounds *in silico* is a complex process. Screening each compound, depending on its structural complexity, requires from a few minutes to hours of computation time on a standard PC. Consequently, screening all compounds in a single database would require years. However, the problem is embarrassingly parallel and the computation time can be reduced very significantly with by distributing data to process over a grid gathering thousands of computers [2,3].

Recently, virtual screening projects on grids have emerged in the perspective to reduce cost and time. They focus on the development of *in silico* docking pipeline on grids of clusters [4,2] but also on the optimization of molecular modelling [5]. Some projects focus on virtual screening deployment on a pervasive grid, or desktops grid, to analyse a specific target [6]. Pharmaceutical laboratories are also interested by the grid concept; Novartis is a successful example of the deployment of an internal grid. They deployed the first automated modelling and docking pipeline on grid. Their vision is to manage the knowledge and the informatics of virtual screening on the grid [7].

Compared to these initiatives, WISDOM is the first attempt to deploy large scale in silico docking on a public grid infrastructure. As highlighted above, previous deployments were either limited to one cluster, to grids of clusters in the tightly protected environment of a pharmaceutical laboratory or to pervasive grids. Compared to pervasive grids, grid infrastructures offer many additional services such as data management, security, grid operation and maintenance.

The potential impact of grids to improve the drug development process is discussed in the next section: more details can be found in the HealthGrid white paper [8].

**1.2. Grid added value for *in silico* drug discovery**

The grid workload and data management systems impact the different steps involved in *in silico* drug discovery. First, the shared resources speed up the execution of time-consuming software like docking and molecular dynamics simulations. Second, the grid helps in identifying new biological targets by offering services such as database replication or workflow management tools. The grid also provides environments and needed services for epidemiology [9], clinical trials and drug delivery monitoring.

A grid infrastructure is very well suited for the collaborative work involved with *in silico* drug discovery researches. Beyond the 24 hours a day availability of resources, the infrastructure perennially gives a stable environment for the scientist for storing results and rerunning experiments. The users' confidence also grows with the improvement of the services security and the quality of user support. The data distribution and management is a crucial point to ease the daily work of *in silico* drug discovery. Finally the philosophy of sharing computing and storage resources in academic projects through the grid opens exciting perspectives for such topics as neglected diseases as it fosters international collaboration especially between western and developing countries.

**1.3. WISDOM, a first virtual screening initiative on grid infrastructures**

WISDOM is an initiative for a World-wide *In Silico* Docking On Malaria. It is the first step toward a virtual screening pipeline on a grid infrastructure. Three goals motivate the initiative. The grid goal is the deployment of a CPU consuming application generating large data flows to test the grid operation and services. Given the very large amount of data involved in the computation, such a large scale deployment is a stressing experiment for the grid infrastructure called a *data challenge*. The bioinformatics goal is the computation of a large scale virtual docking experiment, involving millions of compounds. The biological goal is to propose new inhibitors for a family of proteins produced by *Plasmodium falciparum*.

The paper content is the following: in chapter 2, the three objectives of the data challenge are presented in details. Chapter 3 and 4 propose respectively a brief description of the EGEE infrastructure and of the bioinformatics tools used in WISDOM. Chapter 5 describes in much more details the WISDOM production environment. An analysis of the large scale deployment is proposed in chapter 6 before some conclusions are drawn and perspectives are highlighted in chapter 7.

**2. WISDOM objectives**

**2.1. Grid objective**

A large number of applications is already running on grid infrastructures. Even if many have passed the proof of concept level [10], only few are ready for large-scale production with experimental data. Large Hadron Collider experiments at CERN, like the ATLAS collaboration [11], have been the first to test a large data production system on grid infrastructures [12]. In a similar way, WISDOM aimed at deploying a scalable, CPU consuming application generating large data flows to test the grid infrastructure, operation and services in very stressing conditions.

Docking is – along with BLAST [13] homology searches [14] and some folding algorithms [15,16] – one of the most prominent applications that have successfully been demonstrated on grid testbeds [17]. It is typically an embarrassingly parallel application, with repetitive and independent calculations. Large resources are needed in order to test a family of targets, a significant amount of possible drug candidates and different virtual screening tools with different parameter / scoring settings. This is both a computational and data challenge problem to distribute millions of docking comparisons with millions of small compound files. Moreover, docking is the only application for distributed computing that has prompted the uptake of grid technology in the pharmaceutical industry [18]. The WISDOM scientific results are also a mean of making a demonstration of the EGEE grid computing infrastructure for the end users community, of illustrating the usefulness of a scientifically targeted Virtual Organization, and of fostering an uptake of grid technologies in this scientific area.

**2.2. Bioinformatics objective**

Docking is a first step for *in silico* virtual screening. The goal is to identify which molecules could dock on the protein active sites in order to inhibit its action and therefore interfere with the molecular processes essential for the pathogen. Libraries of compound 3D structures are made openly available by chemistry companies which can produce them.

Many docking software are available either open-source or licensed. WISDOM was the opportunity to compare the results of two software tools on a complete compound database in order to study the correlation between the scores and to analyse possible discrepancies.

## 2.3. Biological objective

Malaria is a dreadful disease affecting 300 million people and killing 1.5 million people every year [19]. Malaria is caused by a protozoan parasite, plasmodium. There are several antimalarial drugs presently available. But the constant emergence of resistance and the costs of the present drugs are worsening the disease condition [19]; therefore it is important to keep exploring new strategies to fight malaria. The one investigated within WISDOM aims at the haemoglobin metabolism, which is one of the key metabolic processes for the survival of the parasite.

There are several proteases involved in human haemoglobin degradation inside the food vacuole of the parasite inside the erythrocytes. Plasmepsin, the aspartic protease of Plasmodium, is responsible for the initial cleavage of human haemoglobin and later followed by other proteases [20]. There are ten different plasmepsins coded by ten different genes in *P. falciparum* (Plm I, II, IV, V, VI, VII, VIII, IX, X and HAP) [21]. High levels of sequence homology are observed between different plasmepsins (65-70%). Simultaneously they share only 35% sequence homology with its nearest human aspartic protease, Cathepsin D4 [22]. This and the presence of accurate X crystallographic data make plasmepsin an ideal target for rational drug design against malaria.

## 3. The EGEE infrastructure

In this chapter, we briefly present the EGEE infrastructure,

The EGEE [23] (Enabling Grid for E-sciencE) project brings together experts from over 27 countries with the common aim of building on recent advances in Grid technology and developing a production Grid infrastructure which is available to scientists 24 hours-a-day. The project aims to provide researchers in academia and industry with access to major computing resources, independent of their geographic location. The EGEE infrastructure is now an operational production grid with a large number of applications installed, exploiting the available resources. The infrastructure involves more than 180 sites spread in Europe, America and Asia.

As stressed above, deployment of docking computations is particularly fitted for grids of clusters like EGEE. The WISDOM application was deployed within the framework of the biomedical Virtual Organization. A Virtual Organization is a grid-wide identification and authorization unit representing a community of users sharing some grid resources. Table 1 gives an overview of the countries with grid sites available for this Virtual Organization, which scaled up to about 2,000 CPUs and 21 TB disk space in the summer of 2005.

| Countries | Number of sites | Countries | Number of sites | Countries | Number of sites |
|---|---|---|---|---|---|
| Bulgaria | 3 | Greece | 3 | Romania | 1 |
| Croatia | 1 | Israel | 1 | Russia | 2 |
| Cyprus | 1 | Italy | 13 | Spain | 7 |
| France | 9 | Netherlands | 2 | Taiwan | 1 |
| Germany | 6 | Poland | 1 | UK | 10 |

Table 1. Countries with grid sites contributing to the EGEE biomedical Virtual Organization during the summer 2005

The LCG-2 [24] middleware deployed on the EGEE infrastructure provides a Workload Management System, a Data Management System, an Information System, an Authorization and Authentication System, an Accounting System, and various monitoring and installation services.

The Workload Management System is responsible for the management and monitoring of jobs submitted from a User Interface. The Job Description Language [25] describes the jobs and their requirements like targeted workstations. A set of services running on the Resource Broker machine matches job requirements expressed in the Job Description Language files to the available resources (as gathered from the Information System), distributes the jobs over the matching Computing Elements, tracks the job status, and allows users to retrieve their job output. Computing Elements are frontals to computing center clusters. They plan jobs execution over the available Worker Nodes through batch schedulers.

**4. WISDOM Bioinformatics tools**

Required inputs to docking software are the three-dimensional structures of the molecules to be docked. WISDOM deployment required selection of a docking software, biological targets and of a library of compounds. For each potentially active chemical compound in the library, the software computes its probability to dock to the target molecule.

The docking tools used in this project are FlexX [26], a commercial software made graciously available by BioSolveIT for a limited time, and Autodock [27], a software which is open-source for academic laboratories and which uses a different docking method. Altogether 4 different parameter sets were generated for FlexX and 2 for Autodock.

The 3D coordinates of three plasmepsin II structures (1lee, 1lf2, 1lf3) and one plasmepsin IV structure (1ls5) were obtained from the Brookhaven protein database [28]. The protein targets were prepared in a format ready to use for Autodock and FlexX.

Chemical compounds were obtained in sybyl mol2 format from the ZINC database [29]. Docking scores of compounds taken from a subset of the ZINC database, the ChemBridge database (~500000 compounds) were computed on the above scenarios using FlexX and Autodock. Another drug like subset (~500000 compounds) of ZINC database was docked on the above scenarios using only FlexX because the computing resources made available to the data challenge were not sufficient to do it also with Autodock. The ChemBridge database molecules were converted from mol2 format to pdbq format in preparation of the Autodock run.

**5. The WISDOM production environment**

The grid is often perceived in life science as a tool for time-consuming applications, but no life science application had been deployed at a large scale on a grid infrastructure before WISDOM. Preparation of the deployment included the development of an environment for job submission and output data collection. This environment had to be able to handle the submission of about 70,000 15-hour long jobs and the collection of the output data. A major issue was to handle job resubmission whenever a job failed for any reason, as the grid success rate was typically of the order of 80 to 85%. Large scale tests were made on the French regional grid Auvergrid to validate the environment and to identify potential issues and bottlenecks.

Other issues were raised by the data challenge, like the usage of licensed software on the grid or the need for a high throughput job submission scheme. In this chapter, we introduce specific issues related to WISDOM deployment. Then, we present the WISDOM production environment which was designed to achieve production of a large amount of data in a limited time with a minimal human cost using EGEE middleware services. We also present the results and lessons learned from the large scale tests deployed on the French regional grid Auvergrid.

**5.1 Specific issues related to WISDOM deployment**

A docking job requires an input file (the target), a database of independent molecules (a compound file), a set of parameters provided in a file or by command line and a docking software. A number of issues need to be addressed to achieve significant acceleration from the grid deployment. Previous experience with LCG middleware indicated potential bottlenecks:

- Grid performances are impacted by the amount of data moved around at job submission. As a consequence, the files providing the 3D structure of targets and compounds should preferably be stored on grid storage elements in preparation for the data challenge.

- The rate at which jobs are submitted to the grid resource brokers must be carefully monitored in order to avoid their overload. The job submission scheme must take into account this present limitation of the EGEE brokering system.
- Grid submission process introduces significant delays for instance at the level of resource brokering. The jobs submitted to the grid computing nodes must be sufficiently long in order to reduce the impact of this middleware overhead.
- Use of licensed software requires designing a strategy to distribute licenses on the grid.

The WISDOM production system was designed to address the issues listed above. It had also to automatically recover on errors to avoid extremely tedious manual intervention. Indeed, the job success rate has kept increasing on the EGEE infrastructure since April 2004 but the achieved efficiency of the order of 85% required handling about 15% of failed jobs.

In the next sections, preparation of the large scale deployment is described as well as the production environment which was developed to maximize job throughput on the grid after analyzing the tests on the Auvergrid regional grid. The strategy adopted for the deployment of licensed software is also discussed.

**5.2 WISDOM preparation**

The preparation took place in several steps.

In order to limit the amount of data transferred at job submission, docking software tools were stored on each Computing Element. The files providing the structures of the target and the compounds were stored on the storage elements. The number of compounds docked per job was estimated so that the job duration was approximately 15 hours. FlexX being much faster than Autodock, different compound files were used as input to the 2 docking software tools. As a result, about 600 MB of data were stored on each grid storage element of the biomedical Virtual Organization.

To achieve this preparation step, two packages were developed. They are responsible for installing the application components on the resources and for testing these components, together with the resources and grid services. Indeed, a stress usage of the grid in a limited time requires resources to be available immediately and reliably. This requires checking the status of the biomedical Virtual Organization services and of each of the grid node committing resources to the Virtual Organization. A very important inefficiency factor comes indeed from sites which are wrongly configured and where jobs systematically fail.

Before WISDOM was launched at its full scale on the EGEE infrastructure in the summer of 2005, deployment went through a ramping process in order to study the performances of the production system and to identify potential bottlenecks. During Christmas break in December 2004, up to 150,000 compounds were docked on the Auvergrid (http://www.auvergrid.fr) infrastructure, a regional grid which gathers the main laboratories of the region Auvergne using EGEE middleware to share technologies, skills and resources. Several instances of variable size were submitted. Table 2 presents some of the parameters which were monitored for 2 of the instances submitted, one of 50 jobs (2,000 dockings) and one of 500 jobs (100,000 dockings):

- The total CPU time corresponds to the cumulated amount of CPU used for a given instance.
- The duration represents the total elapsed time between the submission of the first job and the end of the last job.
- The crunching factor represents the gain of time obtained thank to the grid deployment. It is simply obtained by dividing the total CPU time by the execution duration.
- The grid performance is a measurement of the grid efficiency. It takes into account grid inefficiencies due to job submission failure, aborted jobs, and loss of resources due to competing jobs by other users, etc. It is computed by analyzing all error messages from the grid information system and job log file. For instance, the second test case was launched in a period with an important number of competing jobs, which explains the relatively low grid performance.

- The CPU time for 1 job indicates the average computing time for each job on one of the grid PCs.
- The grid overhead time for 1 job indicates the extra time due to the deployment on the grid. It includes all the extra delays coming from the different grid services (scheduling, queuing...).
- The data transfer time corresponds to the time needed to transfer the input data to the working nodes at job submission time.

| Metrics | 2000 docking in 50 jobs | 100,000 docking in 500 jobs |
| --- | --- | --- |
| Total CPU time | 2.5 days | 188 days (6.3 months) |
| Duration | 2.5 hours | 40 hours |
| Crunching factor | 24 | 113 |
| Grid performance | 50% | 30% |
| CPU time for 1 job | 1.2 hours | 9 hours |
| Grid overhead time for 1 job | 7.2 minutes | 30 minutes |
| Data transfer time for 1 job | < 1 minute | 2,5 minutes |

Table 2: relevant parameters of 2 test cases deployed on the Auvergrid infrastructure during WISDOM preparation.

From table 2, we concluded that submission of longer jobs was definitely more efficient as it reduced the relative grid overhead. Delays due to data transfer were found to be negligible, but this was expected as the different sites used for this test on Auvergrid benefit high bandwidth network connections.

During these tests we also noticed that upgrading grid nodes to new versions of middleware was often generating instability and loss of efficiency. We also understood the necessity to adapt our deployment process to the grid limitations, for instance the limited number of files on each Storage Element or Worker Node or the fact that each Resource Broker uses a different Berkeley Database Information Index for listing available resources. Due to the low BDI update frequency in a large scale system, a resource can still be registered in the information system even if it is in fact out of the grid.

**5.3 WISDOM execution**

Based on the experience acquired during the testing phase on Auvergrid, the WISDOM production system (see figure 1) was developed in Perl, except the multithreaded job submission tool in java. The entry point is a simple command line tool. Its users during the data challenge were members of the Biomedical Task Force which gathers a team of engineers with recognized expertise in application development and deployment. This software environment was developed to allow the submission and monitoring of job sets which were called *instances*. The different jobs of a given instance have the same target input and docking software. They only differ by the molecules of the compound library which are docked. Tasks needed to submit an instance were automatically executed by the WISDOM execution system. The user, authenticated by a proxy certificates, had to start his or her instance execution following a precise submission schedule to avoid to much concurrency between the computation participants leading to a grid overload. Once the computation has started, the WISDOM environment takes care of monitoring jobs and registering results. The user only has to check regularly if the process ran correctly, up to the end of all the jobs belonging to the instance. The overall process progression could be monitored through an output file for follow-up messages and an error file in case of problem.

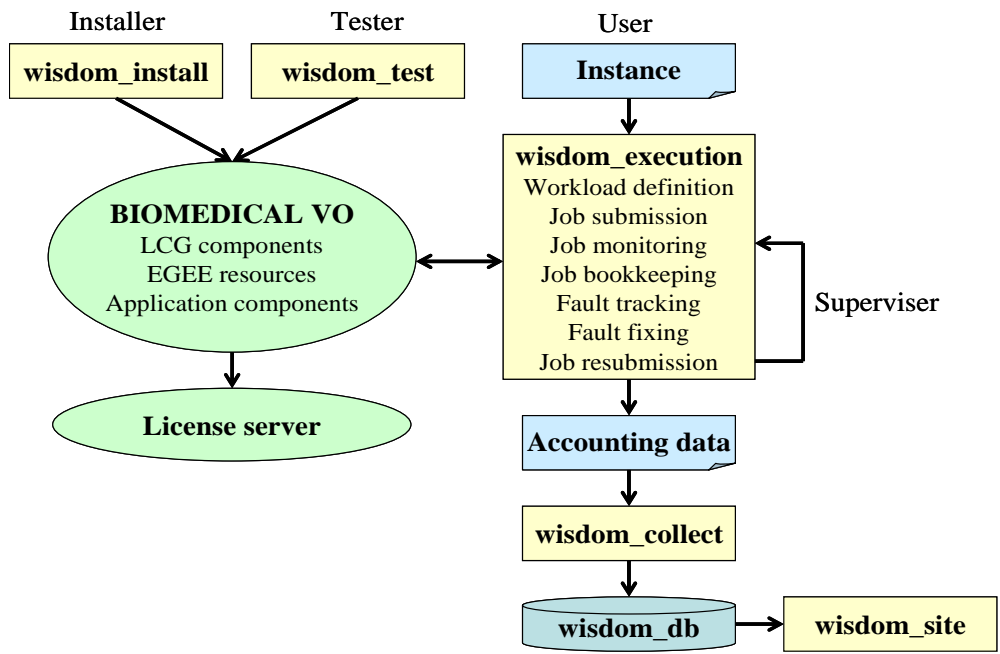

Figure 1: Design of the WISDOM production system.

For each instance, a configuration file contained the instance information (software, target, database, parameter settings) and the grid parameters (number of jobs for the instance, Resource Brokers, Computing Elements, Storage Elements). A shell script and a Job Description Language file were created for each job and used by the submission tool.

On the Worker Node, after the environment was configured, the shell script downloaded the database file from a Storage Element chosen by the Information System using the LCG API. Then binaries were called with the target and parameter settings transferred with the job. The compressed result was stored on a Storage Element and registered in the grid file catalogue. A backup copy was also generated on another Storage Element. For sake of simplification, the most relevant metadata relative to the output (software, parameter settings, compounds database, etc.) were stored in the name of the output itself. Output, errors and accounting messages were transferred on the User Interface.

A multithreaded job submission was developed specifically for a bulk and efficient submission on the Resource Brokers to address the limitation of LCG API which submits jobs one by one on one Resource Broker (with a minimal latency time of 7 seconds). The aim of the tool was to reduce time needed for sequential submission of the jobs on Resource Brokers by parallelising jobs submission on several Resource Brokers.

Once the jobs were submitted, the supervisor needed a real-time overview of the production. For a large number of jobs, the WISDOM environment includes an automatic monitoring tool which was developed to check the status of the jobs with the bulk monitoring LCG API and to react if needed.

Discovering and fixing failures was crucial for the process, and for the evaluation of grid performances. For each job that failed, the system was able to track the fault in the Workload Management System status message or in the job output content. After any correcting operation, like deletion of a faulty Computing Element in the attributes of the Job Description Language file, the job was resubmitted. Preventive actions were also taken during the process. Then GridFTP was used to transfer data on the grid automatically from the Worker Node when the LCG Data Management Service failed. The output files were also carefully stored on several Storage Elements for security concerns.

**5.4. Licenses management**

During the data challenge, commercial docking software (FlexX) with floating licenses was used on the EGEE infrastructure. 3000 floating licenses were made available by BioSolveIT during 3 weeks to be distributed on the grid.

Each job using FlexX software was contacting the Flexlm server at the beginning of the job and asked for a license, namely an ASCII file with specific keys generated for this server. Then the job was able to run without connection to the license server. Accessing floating licenses on the grid behind firewalls required known IP and the opening of two specific ports for institutes hosting Worker Nodes. During the data challenge first 3 weeks, FlexX licenses were available through the license server. Afterwards, the Autodock phase started in August for about 2.5 weeks.

## 6. WISDOM deployment

The deployment took place in July and August 2005. During this period, 10 users have been launching jobs from 5 User Interfaces, monitoring the process with the help of the WISDOM environment and interacting with the user support of the EGEE project and the nodes administrators. 72,751 jobs were launched for a total of 80 CPU years, producing 1 TB of data (500 GB, doubled for the back-up).

In this section, we are going to further document the deployment. We will first describe achievements in terms of scale, and then we will discuss the grid performances measured during the deployment. Finally, the performances of the different grid services are also discussed.

### 6.1. Achieved deployment

For a sake of simplicity, the WISDOM deployment has been split into 6 phases:

- Phase n°1 corresponds to the high throughput submission of FlexX jobs against the ChemBridge database of compounds. Many problems due to the grid (sites, Resource Broker…) and to WISDOM production system (load balancing, process management…) were discovered during this early period.
- Phase n°2 corresponds to the resubmission of failed jobs after phase 1
- Phase n°3 corresponds to a second high throughput submission of FlexX jobs with the second drug like compound base. This phase ended August, the $1^{st}$, when the number of available FlexX licenses was reduced to 100.
- Phase n°4 corresponds to the resubmission of failed jobs after phase 3
- Phase n°5 corresponds to a first high throughput submission of Autodock jobs with ChemBridge database of compounds

- Phase n°6 corresponds to the resubmission of failed jobs after phase 5.

Figure 2 shows the number of docked compounds over time during the months of July and August 2005. Figure 3 shows the number of running and waiting jobs vs. time. Figure 4 shows the amount of transferred output vs. time.

The different phases show very different patterns:

- Phases 2, 4 and 6 are resubmission phases where the number of jobs submitted and the amount of data produced are significantly lower than in the high throughput docking phases 1, 3 and 5
- Phase 1 corresponds to a ramp-up phase where many bugs were identified and addressed
- Data output as well as docking throughput were highest in phase 3 of production with FlexX. Indeed, Autodock software is about 3 times slower than FlexX software.

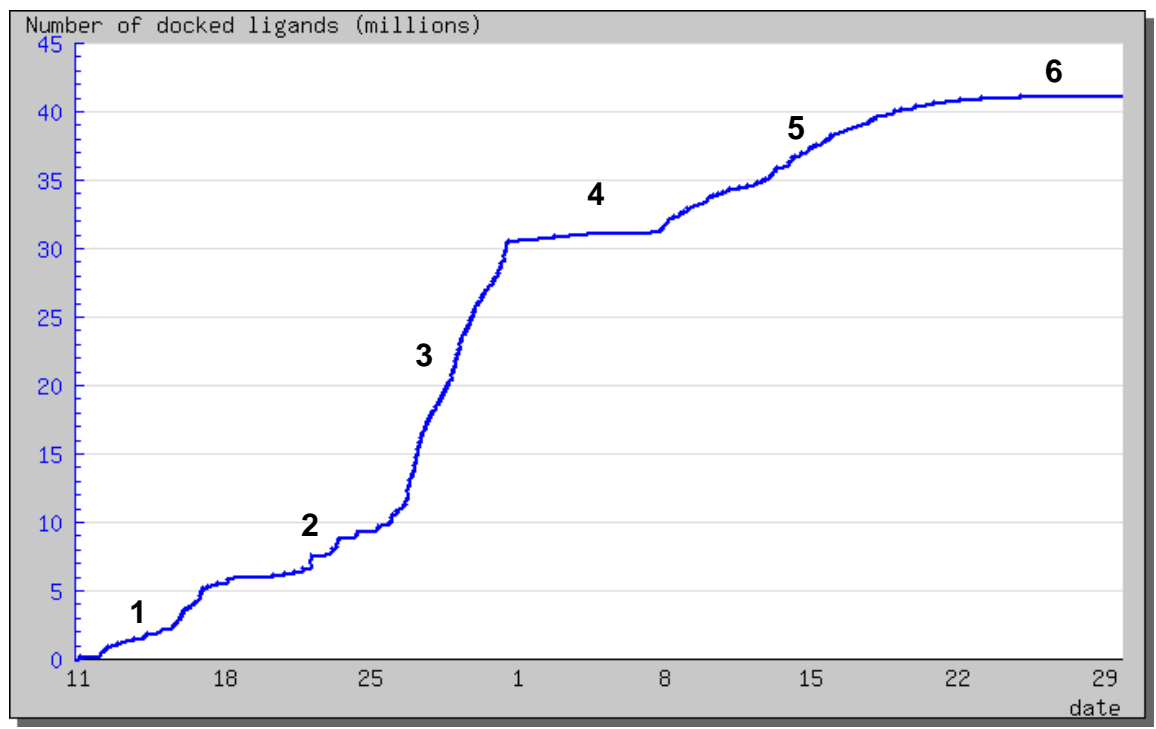

Figure 2: Number of docked compounds vs time.

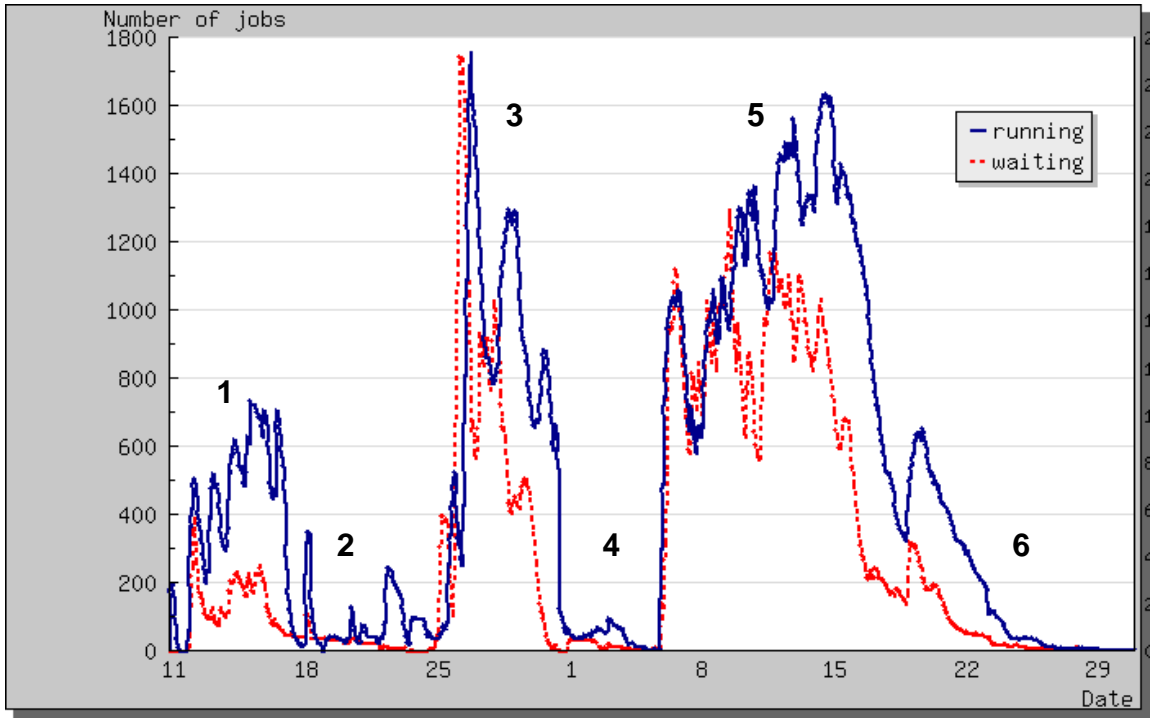

Figure 3: Number of running and waiting jobs vs. time.

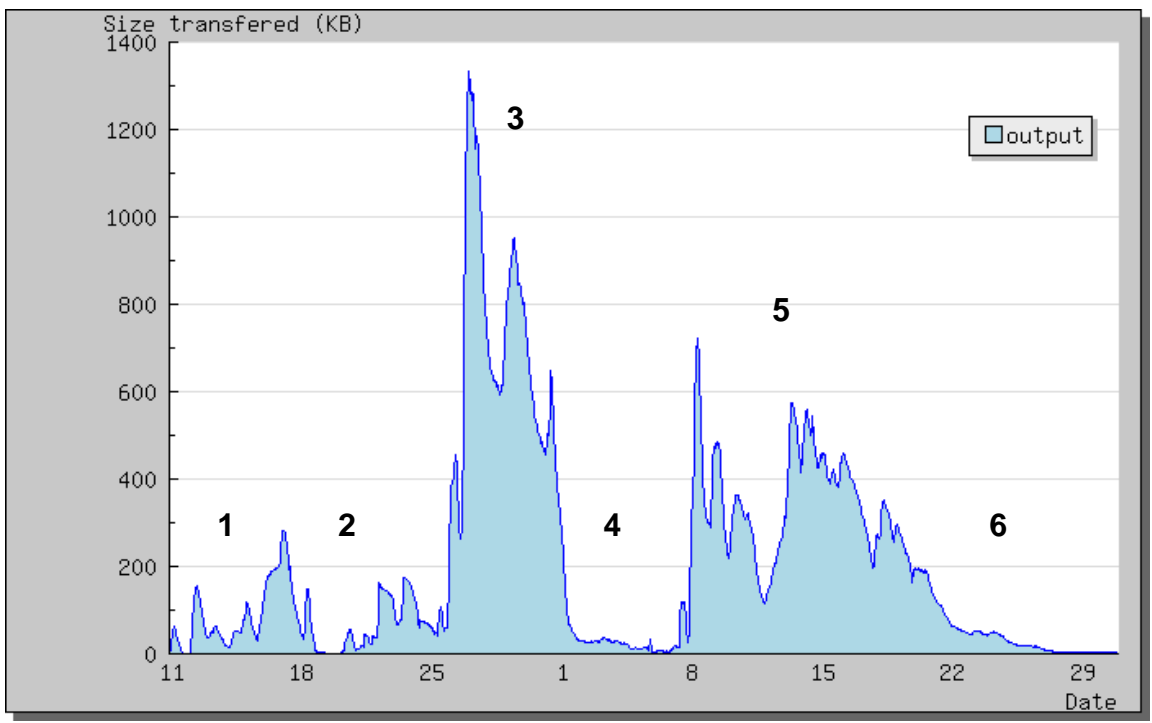

Figure 4: Amount of transferred output vs. time.

Shift between the curves of waiting and running time on figure 3 illustrates the latency introduced by the grid. This latency is further documented in the "grid node performances" section but it is worth noticing already that such latency, of the order of a few hours, is acceptable only if the submitted jobs are themselves hour-long jobs.

In the following, the phases 1 to 4 will also be called the FlexX phase while the last 2 phases will also be called the Autodock phase.

| Metrics | Total | FlexX phase | Autodock phase |
| --- | --- | --- | --- |
| Cumulated number of docked compounds (in millions) | 41,27 | 31,41 | 9,87 |
| Effective duration | 37 days | 22 days | 15 days |
| Number of docked compounds / hour | 46475 | 59488 | 27417 |
| Crunching factor | 662 | 411 | 1031 |
| Number of jobs submitted | 72751 | 41520 | 31231 |
| Number of grid Computing Elements used | 58 | 56 | 57 |
| Number of Resource Brokers used | 12 | 12 | 11 |
| Maximum number of jobs running in parallel on the grid | 1643 | 1008 | 1643 |
| Volume of output data | 946 GB | 506 GB | 440 GB |
| Total CPU time | 80 years | 29,5 years | 50,5 years |
| Effective CPU time used by success jobs | 67,2 years | 24,8 years | 42,4 years |
| Overhead time | 77,1 years | 25,9 years | 51,2 years |

Table 3: Performance metrics of the WISDOM deployment.

Table 3 presents several parameters relevant to evaluate WISDOM deployment scale for the Autodock and FlexX phases. The following points are worth mentioning:

- The number of docked compounds is a critical parameter for computational chemists. In only 37 days of effective computing, 41.27 millions number compounds were docked. As stated above, FlexX is significantly faster than Autodock and the FlexX phase allowed computing three times more compounds than the Autodock one.

- The number of docked compounds per hour is not a factor 3 larger for the FlexX phase compared to the Autodock phase because the number of FlexX licenses was limited to 1,000 while the number of CPU used during the Autodock phase reached more than 1,500 nodes.

- The average crunching factor is 662, and reached 1031 during the Autodock phase.

We are going to detail in the next section the inefficiencies that generated the overhead time.

**6.2. Grid node performances**

Figure 5 illustrates the different delays introduced by the grid deployment for 4 sites of the biomedical Virtual Organization used for WISDOM. Jobs undergo different states when submitted to a grid Computing Element:

- "Submitted" corresponds to jobs submitted by the user through the User Interface and not yet handled by the Resource Broker. It corresponds also to jobs failed and automatically resubmitted by the Resource Broker.

- "Waiting" corresponds to jobs accepted by the Resource Broker but which are not yet allocated to a Computing Element

- "Ready" corresponds to jobs for which the matching resources are found and which are submitted to a Computing Element

- "Scheduled" corresponds to jobs accepted by a Computing Element and which are queuing for execution

- "Running" corresponds to jobs executed on a Worker Node.

A grid node performance depends of several parameters like scheduling policies, Worker Node configuration and system failures. Figure 5 shows the average time spent by the jobs submitted to the 4 sites in the different states.

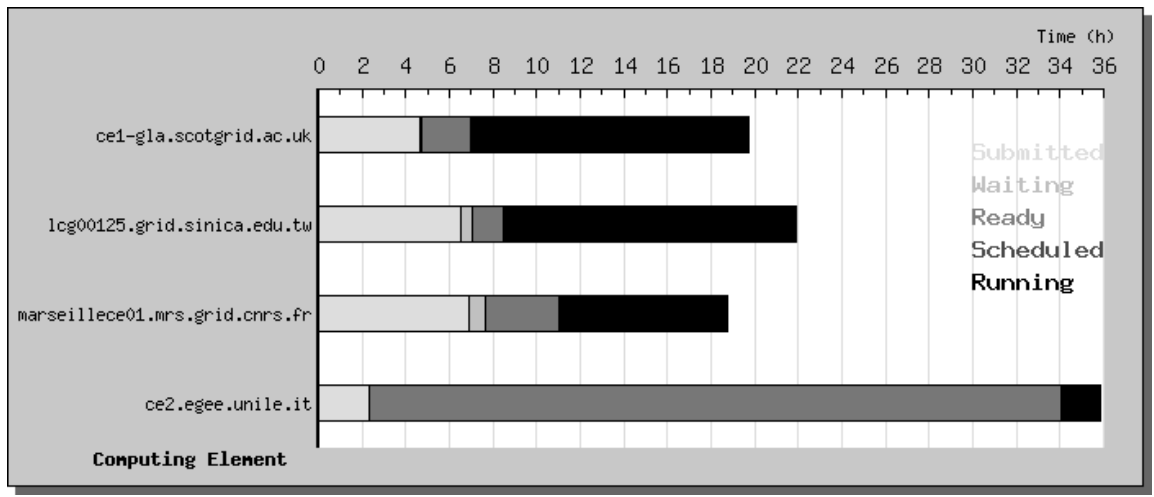

Figure 5: Average time for the different status of a job on a computing element.

Delays introduced by the grid have different explanations for the 4 sites presented on these figures:

- Scotgrid.ac.uk is a node with about 200 Worker Nodes that did not experience break down during the data challenge. The very short time spend as "scheduled" shows that this node was not busy with concurrent jobs during the data challenge. The time spent at submission, corresponding to the time needed to access the Resource Broker, is more important.
- Sinica.edu.tw is a node similar in size to scotgrid.ac.uk except its location in Asia. Additional time spent as "submitted" can be attributed to the communication time between European Resource Brokers and Taiwan.
- The third grid node located in Marseille hosts 30 Worker Nodes. It is interesting to notice that the time spent in the "submitted" state is comparable to what is observed in Taiwan. This is due to the fact that the node limited the number of scheduled jobs to 30. This limitation was ignored by the Resource Broker and therefore, the extra jobs submitted to this site were discarded immediatly. This example shows how a node can have a special configuration not taken into account by the middleware.

- The last node shown on the figure is unile.it with only few Worker Nodes. Such a small node should not receive a large number of jobs. But a failure of the grid Information System directed almost 100 jobs to the node which increased very significantly the scheduled time.

## 6.3. Analysis of job success rate

The success rate definition used in this paper is the same as the ratio used for measuring EGEE Quality of Service. The formula stands as follows:

Successful jobs / (submitted jobs – cancelled jobs)

where successful jobs are jobs which have been executed successfully, submitted jobs are the jobs launched by the user from the User Interface, cancelled jobs are jobs cancelled by the user.

The proposed definition of the success rate is not completely relevant from a user point of view as a successful job as seen from the grid can be unsuccessful from a user perspective if it did not produce the expected output data.

Finally, some of the jobs have to be cancelled for reasons external to the grid, for instance failures of the WISDOM execution environment or break downs of the FlexX license server. Table 4 shows the success rates during the data challenge and more specifically in its two specific phases.

| Metrics | Total | FlexX phase | Autodock phase |
| --- | --- | --- | --- |
| EGEE success rate | 77 % | 80,4 % | 71,9 % |
| Success rate after checking output data | 46,2 % | 33,8 % | 64,4 % |
| Success rate after checking output data and subtracting WISDOM and server license failures | 63 % | 61,6 % | 65 % |

Table 4: Efficiency metrics of the WISDOM deployment.

An analysis of the origin of failures is summarized on table 5 with their corresponding rates.

During the FlexX phase, the dominant source of failures was the license server which was distributing tokens to all the jobs running on the grid. This bottleneck can be overcome by having several license servers available.

|  | Rate | Reasons |
|---|---|---|
| Success rate after checking output data | 46 % |  |
| Workload Management failure | 10 % | Overload, Disk failure |
|  |  | Mis-configuration, disk space problem |
|  |  | Air-conditioning, electrical cut |
| Data Management failure | 4 % | Network / connection |
|  |  | Electrical cut |
|  |  | Other unknown causes |
| Sites failure | 9 % | Mis-configuration, Tar command, disk space |
|  |  | Information system update |
|  |  | Job number limitation in the waiting queue |
|  |  | Air-conditioning, Electrical cut |
| Unclassified | 4 % | Lost jobs |
|  |  | Other unknown causes |
| Server license failure | 23 % | Server failure |
|  |  | Electric cut |
|  |  | Server stop |
| WISDOM failure | 4 % | Job distribution |
|  |  | Human error |
|  |  | Script failure |

Table 5: Origin of failures during the WISDOM deployment with their corresponding rates.

The second major source of failure were workload management and site failures, including overload, disk failure, node mis-configuration, disk space problem, air-conditioning and electrical cut. To improve the job submission process, an automatic resubmission of jobs was included in the WISDOM execution environment. However, the consequence of automatic resubmission was the creation of several "sink-hole" effects where all the jobs are attracted to a single node. These sink-hole effects were observed when the status of a Computing Element was not correctly described in the information system. If a Computing Element already loaded is still viewed as completely free by the Information System, it keeps receiving jobs from the Resource Broker. If the Computing Element gets down, all jobs are aborted. If the Computing Element can support the excessive number of jobs, the processing time is going to be very long.

Most of Data Management System failures were circumvented by the back-up system. Finally, unclassified failures account for 4% inefficiency. This illustrates the work which is still needed to improve grid monitoring.

Other large scale deployments on EGEE report similar success rate [12] although the reported causes of job failures are partially different.

**6.4. Analysis of grid services**

The WISDOM data challenge was the opportunity to test the different grid components within the framework of a large scale deployment. Each service has an important role in the process. In this section, we describe the main issues encountered with the different grid services.

**6.4.1. Information system**

To process a job, a Resource Broker chooses the best Computing Element using the information provided by the Information System. We identified the following issues in relation to EGEE Information System:

- a Computing Element can have a policy unknown to the Information System like a limitation on the number of authorized biomedical jobs.
- When the Grid Information Index Service of a Computing Element is down or very slow, it provides obsolete information on its status. This is misleading for the Resource Broker which may keep allocating jobs to this Computing Element although it is saturated.
- The Information System is updated every 2 minutes. This has to be compared to the normal submission time to a Resource Broker which is 7 seconds. As a consequence, when jobs are simultaneously submitted to several Resource Brokers, these services have exactly the same image of the system and therefore distribute jobs in a very similar manner. This generates overload of the best ranked sites.

In summary, our experience shows that detailed information on Computing Element configuration must be known to the Information System in order to manage large scale submission of jobs. Because of the limitations identified, a strategy was adopted not to submit jobs in bursts but rather to have a constant submission flow to limit the impact of missing information and sink-hole effects.

**6.4.2. Workload management system**

The Resource Broker allocates jobs to a Computing Element depending of the job attributes described using the Job Description Language. Ranking of the Computing Elements depends on several criteria including the number of free Worker Nodes or the number of scheduled or running jobs.

Resource Brokers turn out to be significant bottlenecks of the system. Several failures of the most used Resource Brokers were observed including disk crashes or overloaded services. As a consequence, hundreds of jobs were lost or retrieved with difficulty. In practice, the measured time for job submission was closer to 30 seconds than 7 seconds. WISDOM users had to change regularly Resource Brokers to limit their workload and unstability. They were often left with the task to allocate jobs to Computing Elements by hand.

Our experience shows that there is a real need for synchronization between Resource Brokers to be able to send a job through a given Resource Broker and check its status or retrieve results via another one. It means that job databases should not be stored on the Resource Brokers but somewhere on the grid where it can be reached by any Resource Broker and possibly replicated. The idea would be to have a job management similar to the data management, with the notion of Logical File Name for a job.

On such large scale experiments, bulk submission, grouped monitoring, and partial error mechanisms are critical. On the EGEE infrastructure, bulk sumbission and grouped monitoring are implemented to some extent, although the application needs to limit the stressing level on the workload management services. Partial error recovery is still handle at the level of the WISDOM environment.

### 6.4.3. Data Management system

The Data Management System certifies data copy, registration and replication. This service is also a potential bottleneck, because there is only one file catalogue per Virtual Organization. Main cause of failure for the previous ATLAS Rome Production Experiment [12], the EGEE Data Management System was significantly improved for the WISDOM data challenge. In case of breakdown of the Data Management System, the back-up solution for data transfer was to use GridFTP retry system for the input copy from a Storage Element to a Computing Element and for the output copy, registration and replication from a Computing Element on two Storage Elements. But failures like an electric cut of the Computing Center of Lyon, IN2P3, hosting the unique file catalogue, proved the limitation of the system.

The Data Management System could be improved by replicating the service and file information on several places. An automatic retry after checking the size and the integrity of the copied data could be developed similarity to the retry system for job failures.

**6.4.4. Resource centres**

The resource centres, i.e. the Computing Elements, the Worker Nodes and the Storage Elements, were essential to the data challenge success. The 3 Storage Elements storing the output data on disk or tape worked very well despite the concurrent accesses. Difficulties encountered using Computing Elements have been discussed previously.

The need for regularly testing all the resources of the biomedical Virtual Organization emerged during the preparation of the large scale deployment. Today, automatic testing is only performed on the EGEE infrastructure for a special test Virtual Organization. These tests allow to automatically extract failing resources from the production grid.

**6.4.5. User interface**

The User Interface is the entry door where the user is authenticated and authorised to submit jobs and manage data. Its disk capacity must be sufficient to receive all the information during and after the execution. Preliminary tests are needed on each User Interface to evaluate its capacity to control a large number of jobs with EGEE APIs. Java and Perl are also required.

**6.4.6. License server**

The license server aims at authorising the use of commercial software for a large number of jobs. Only few institutes were not able to open the necessary ports for security reasons. They did not participate to the FlexX part of the data challenge. The server failed a few times because of a limited number of file descriptors on the system which prevented new socket connections. It was necessary to increase this number to allow new jobs to get a license (the number of possible socket connections must be at least equal to the number of available licenses). In the future, an interesting improvement would be to integrate the server into the grid, so that it can check the certificates of the users requiring a license.

### 6.4.7. Error messages

When a job is aborted, a status message is sent to help understanding and solving the problem. There is a list of possible reasons for job abortion, but the message is often not clear and not sufficient to know exactly what happened and what the source of the error is. The help from user support and site administrators is therefore crucial.

### 7. Conclusion and perspectives

The WISDOM data challenge is the first large deployment of a biomedical application on a grid infrastructure. On a biological side, the aim of the application was to identify new inhibitors for a family of proteins produced by *Plasmodium falciparum* through *in silico* virtual docking on a grid infrastructure. This paper has described how this application allowed to test the grid operation and services for a CPU consuming application generating large data flows.

From 11 July 2005 until 19 August 2005. up to 1700 computers were simultaneously used in 15 countries around the world to dock over 41 million compounds. On the biological side, the data challenge produced a large amount of output for analysis. Results extracted 10% of the compounds with key interactions and good scoring. Top scoring compounds possess basic chemical groups like thiourea, guanidino or amino-acrolein core structure.

The data challenge has also been a very useful experience to identify the limitations and bottlenecks of the EGEE infrastructure. The WISDOM production system developed to submit the jobs on the grid accounted for a small fraction of the failures as well as the grid management system. On the other hand, the resource brokers have been significantly limiting the rate at which the jobs could be submitted. Another significant source of inefficiency came from the difficulty for the grid information system to provide all the relevant information to the resource brokers when they distribute the jobs on the grid. As a consequence, job scheduling was a time-consuming task for the WISDOM users during all the data challenge due to the encountered limitations of the information system, the computing elements and the resource brokers.

Finally, the necessity to deploy licensed software during one of the deployment phases has generated a single point of failure ignored by the information system. The development of a grid service to manage license software is under way to address this limitation.

Following the work presented in this paper, a new data challenge dedicated to *in silico* docking against avian flu was deployed on the EGEE and Auvergrid infrastructures during the spring of 2006. This new deployment required a much shorter preparation, about a month, and took advantage of the experience acquired with WISDOM. Another large scale deployment focussed on neglected diseases is foreseen for the fall of 2006.

**Acknowledgements**

EGEE is a project funded by the European Union under contract INFSO-RI-508833. Auvergrid is a project funded by Conseil Régional d'Auvergne. The authors thank particularly the RUGBI (Réseau Gen'Homme) and GLOP (ACI GRID) projects, funded by the French Ministry of Research, the Biomed Task Force members, the EIS team, the JRA2 members, Lydia Maigne and Cheick O. Thiam. The authors acknowledge the support of BiosolveIT which made freely available up to 1000 FlexX licenses.

The following institutes contributed computing resources to the data challenge: ASCC (Taipei); IPP-BAS, IMBM-BAS and IPP-ISTF (Bulgaria); CYFRONET (Poland); ICI (Romania); CEA-DAPNIA, CGG, IN2P3-CC, IN2P3-LAL, IN2P3-LAPP and IN2P3-LPC (France); SCAI (Germany); INFN (Italy); NIKHEF, SARA and Virtual Laboratory for e-Science (Netherlands); IMPB RAS (Russia); UCY (Cyprus); AUTH FORTH-ICS and HELLASGRID (Greece); RBI (Croatia); TAU (Israel); CESGA, CIEMAT, CNB-UAM, IFCA, INTA, PIC and UPV-GryCAP (Spain); BHAM, University of Bristol, IC, Lancaster University, MANHEP, University of Oxford, RAL and University of Glasgow (United Kingdom).

**Legends**

Figure 1: Design of the WISDOM production system.

Figure 2: Number of docked compounds vs time.

Figure 3: Number of running and waiting jobs vs. time.

Figure 4: Amount of transferred output vs. time.

Figure 5: Average time for the different status of a job on a computing element.

**Captions**

Table 1: Countries with grid sites contributing to the EGEE biomedical Virtual Organization during the summer 2005.

Table 2: Metrics of 2 test cases in the deployment preparation on the Auvergrid infrastructure.

Table 3: Performance metrics of the WISDOM deployment.

Table 4: Efficiency metrics of the WISDOM deployment.

Table 5: Origin of failures during the WISDOM deployment with their corresponding rates.